\documentclass{aa} 
\usepackage{float} 
\usepackage{graphicx}  
\usepackage{graphics}
\usepackage{txfonts}  
\usepackage{natbib}
\usepackage{float}
\usepackage{lscape}  
\usepackage{wasysym}
\bibpunct{(}{)}{;}{a}{}{,}
\usepackage[dvipsnames]{xcolor}
\usepackage[normalem]{ulem}

\begin{document}  

\title {Matter accretion in metal-poor stars down to extremely metal-poor stars and the lithium problem}
               
\author{M. Deal\inst{1}  \and O. Richard \inst{2} \and S Vauclair\inst{3,4}}
  
\institute{Instituto de Astrofísica e Ciências do Espaço, Universidade do Porto, CAUP, Rua das Estrelas, PT4150-762 Porto, Portugal 
          \and
Laboratoire Univers et Particules de Montpellier, Université de Montpellier, CNRS, Place Eugène Bataillon, 34095 Montpellier, France  
          \and
Université de Toulouse; UPS-OMP; IRAP; Toulouse, France
          \and
CNRS; IRAP; 14 av. Edouard Belin, F-31400 Toulouse, France}

\date{\today}

\abstract
{The formation and evolution of light elements in the Universe act as important cosmological constraints. It has long been assumed that the oldest stars of the Galaxy  display the primordial lithium abundance in their outer layers, although studies of stellar physics have proven that this abundance must have decreased with time. The primordial Li abundance deduced from the observations of the cosmic background is, indeed, larger than the maximum observed in these stars. Recent observations have given evidence of a large Li abundance dispersion in very metal-poor stars. Many of these stars are carbon-rich, that is, the so-called carbon-enhanced metal-poor (CEMP) stars. A large number of them also present overabundances of s process elements (CEMP-s).}
{We address the general question of the observed abundances in metal-poor stars and we focus our study on the case of CEMP-s stars. We study how the accretion of the wind of stellar companions, especially asymptotic giant branch (AGB) stars, modifies the element abundances of metal-poor stars and, in particular, lithium, taking into account the stellar structure and the hydrodynamic processes that take place after accretion. We compare the results with the observations of lithium and heavier elements in these old stars on the main sequence.}
{We use the Montr\'eal/Montpellier stellar evolution code, which includes atomic diffusion and thermohaline convection, to compute the internal structure of the proto-CEMP-s stars and their evolution, from [Fe/H]$~=-2.31$ down to [Fe/H]$~=-5.45$. We study a number of cases that vary according to the masses of the stars, their ages, metallicities, and the distances to their respective companions.}
{We show that the observations of lithium dispersion that is associated (or not) with carbon enrichment are well accounted for in terms of accretion on to the metal-poor stars of the winds of stellar companions, with accreted masses smaller than those considered in previous studies. The derived primordial value is in accordance with the cosmological results.
}
{}

  
\titlerunning{}
  
\authorrunning{}  

\maketitle 

\section{Introduction}
\label{sec:intro}
Most of the light elements that are currently observed in the Universe, namely, D, $^3$He, $^4$He, and $^7$Li, were produced about five minutes after the Big Bang as part of the so-called Big Bang nucleosynthesis (BBN). The precise quantity of the elements formed in this process closely depends on the baryonic density of the Universe $\Omega_B$ (ratio of the baryonic to the total matter-energy density). For this reason, determining their original abundances from the observed ones by taking into account their abundance evolution over time is crucial from a cosmological viewpoint. The largest uncertainties in this respect lie in the way their time evolution is evaluated. Comparisons between the constraints obtained on $\Omega_\mathrm{B}$ from BBN and, independently, from the cosmological background lead to important information on the mass-energy content of the Universe.

Each of the four primordial isotopes represents a special case. The determination of their abundances at the present time and the evaluation of their evolution with time since the Big Bang are different. Here we focus on the lithium problem. The main constraints on this element rely on spectroscopic observations of the oldest stars of the Galaxy. Lithium is observed through only one doublet, at 607.78nm. When \cite{spite82} discovered that this lithium feature was similar in all the PopII stars that they observed, they deduced that they had reached the primordial lithium abundance, and they gave the value of $A(Li)=2.05\pm{0.15}$ in the usual scale\footnote{A(Li)=$\log(N_\mathrm{Li}/N_\mathrm{H})+12$}. Since this work, many other studies found that the dispersion in this lithium plateau was small and within the error bars \citep{rebolo88}.

Meanwhile, stellar physicists immediately pointed out that the currently observed lithium abundance in these stars could not, in fact, be the original one \citep{michaud84}. The lithium nucleus is easily destroyed by proton capture at a temperature of about 2.5 million degrees, not very deep inside the star. If the stellar gas is stable against turbulence or any kind of mixing, the lithium must have partially sunk from the outer layers owing to atomic diffusion (dominated by the gravitational settling contribution). In the case that some mixing exists inside the star that is strong enough to prevent this settling, then the lithium atoms are transported down to the layers where they are then destroyed. This was confirmed by several subsequent works from various studies, showing that the original lithium abundance in these stars should have been at least twice that which had been observed \citep[e.g][]{vauclair88}. More recently, \cite{korn06,Kornetal2007} studied the lithium and heavy element abundances in the globular cluster NGC 6397. They could reproduce both the lithium plateau and the iron abundance observed in the stars of this cluster, with an original lithium abundance two times the present one, as well as depletion processes including both atomic diffusion and parametrised turbulence. Similar results were found in the globular clusters NGC 6752 \citep{Gruytersetal2013} and M30 \citep{Gruytersetal2016}.

The basic reasons supporting the belief that the observed plateau abundance was primordial were twofold. The first reason was the very small dispersion, which seemed, at first sight, difficult  to reconcile with any type of depletion. Precise computations show, however, that the mass range of the plateau stars is small, between $0.70$ and $0.78$ $M_\odot$, and their internal structures do not differ very much, with metallicities between [Fe/H] = $-1.31$ and $-2.31$, so that transport processes lead to similar depletion factors \citep[see Figure 2 of][]{richard02}. The second reason was the peculiar behaviour on the part of the observed lithium abundance with stellar metallicity. In young metal-rich stars, the lithium dispersion is large. The maximum lithium abundance is about ten times that which is observed in old stars and the dispersion is the signature of lithium depletion in these stars via settling or nuclear destruction. On the other hand, when observing older stars with smaller metallicities, it appeared that the dispersion would decrease rapidly, leading to the plateau value \citep{steigman02}. We notice, however, that lithium was not detected in a few of these stars \citep{spite93}, leading to upper limits that were not taken into account in most subsequent studies.

More recently, two different observations have thrown stones into the water. Firstly, there were the observations of the cosmic microwave background (CMB) by   the Wilkinson Microwave Anisotropy Probe (WMAP) and then Planck proved that the primordial lithium abundance is indeed larger than the plateau value \citep{Cyetal2016,coc17}. Secondly, the spectroscopic observations of very low-metallicity stars show that, contrary to expectations, the lithium dispersion is very large \citep{bonifacio07,cayrel08,sbordone10}.

The extremely metal-poor stars (EMP) are classified in various types, according to their observed chemical composition \citep{BeCh2005}. A large number of these stars are carbon-rich, namely, carbon-enhanced metal-poor stars (CEMP) and some of them are also enhanced by s-process elements (CEMP-s).

The formation of CEMP-s stars is currently attributed to the accretion of the wind of an  asymptotic giant branch (AGB) companion \citep{mcclure90,jorissen16}. In this case, the accreted matter is diluted inside the newly formed CEMP-s star due to macroscopic and microscopic transport processes. The study of these transport processes is necessary for improving our understanding of the surface abundance patterns of CEMP-s stars. One of these processes is thermohaline convection. \cite{stancliffe07} presented the first models of CEMP-s star formation that include this process. Then atomic diffusion without radiative acceleration, in the trace element approximation, was included in the models \citep{stancliffe08,stancliffe09}. More recently, the effects of radiative accelerations (without rotation) and the effects of rotation (without radiative accelerations) were included in the models \citep{MaSt2016,MaSt2017}. All these studies showed that these processes are crucial for explaining the formation of CEMP-s stars and are able to reproduce the surface abundances of these stars relatively well from the main sequence to the red giant phase at [Fe/H]$=-2.31$.

In this paper, we  simultaneously treat two questions, namely: lithium depletion from the primordial value to the present one observed as an upper envelope in the old stars and the reason for the increase of the dispersion at very low metallicity. The CEMP-s stars offer the opportunity to address these questions in the presence of accretion, which has a strong impact on lithium abundance, particularly because of thermohaline convection. \cite{stancliffe09} already addressed the lithium abundance in CEMP-s stars in this context for stars with [Fe/H]$=-2.31$ and accretion masses from 0.001 to 0.1~$M_\odot$ according to the Bondi--Hoyle--Lyttleton theory \citep{bondi44,hoyle39}. They found that lithium was completely depleted at the surface of their models (in contradiction with the lithium determined from observations), and they concluded that either the efficiency of thermohaline was overestimated or the accreted masses were too large. 

In this study, we focus on main-sequence CEMP-s stars, where the surface abundance pattern is not affected by convective dredge-up, in order to study, specifically, the effect of atomic diffusion (including radiative accelerations), turbulent mixing, and thermohaline convection on lithium surface abundances. We computed models with the Montreal/Montpellier code. This code uses completely different numerical schemes, micro-physics, and physical prescriptions than were applied in previous studies. This allows for comparisons and a verification of the robustness of the results. Furthermore, we explore the effect of smaller accreted masses and we also extend the study to lower initial metallicities.

In Section 2, we analyse the status of the observed lithium in CEMP stars. In Section 3, we describe the stellar code that we use in the computations, the input physics, the turbulent mixing prescription, and the processes of lithium diffusion and destruction. Section 4 is devoted to the consequences of the accretion of AGB winds on extremely metal-poor stars according to the stellar masses, the chemical composition of the wind, and the age of the EMP star at the time of accretion. We also discuss thermohaline convection, which appears as soon as heavy matter is accreted at the stellar surface. The quantitative results of the accretion on the lithium abundances are given in Section 5 for a large number of different cases. The scenario for heavy elements is also discussed in this section, as well as the carbon isotopic ratios. Section 6 is devoted to a general discussion of these results, along with a qualitative discussion of cases of various observed EMP stars showing different chemical characteristics in comparison to the CEMP-s stars.

\section{Lithium in CEMP stars}

The observations of the most metal-poor stars of the Galaxy show specific patterns in heavy elements that have to be taken into account when discussing the lithium abundances. Most of the observed EMP stars present an overabundance of carbon and are therefore classified as CEMP stars. The exact proportion of these stars is however difficult to evaluate. We must start by making note of the two different definitions for CEMP stars published in the literature. Firstly, \cite{BeCh2005} take CEMP to refer to all stars with [C/Fe] > 0.7, whereas \cite{aoki07} put the limit up to [C/Fe] > 1. In addition, it is not always possible to measure the carbon abundance in these stars owing to low signal-to-noise ratios (S/N) in the stellar spectra \citep{bonifacio15,aguado19}. In spite of these difficulties, it appears to be well-established that the carbon enhancement increases with decreasing metallicity \citep{BeCh2005,bonifacio15,chiaki19}.
 \cite{aguado19} report a fraction of CEMP stars (with [C/Fe] $\ge +0.7$) of $41 \pm 4\%$ for stars with $-3 <$ [Fe/H] $< -2$ and $58 \pm 14\%$ for stars with [Fe/H] $< -3$.

Some of the CEMP stars, called CEMP-s, are also enriched in s-process elements, whereas some others, the CEMP-rs are overabundant in both r and s elements. Very few CEMP stars are overabundant in r-process elements only. There are also CEMP stars with a normal (solar) pattern of heavy elements, which are referred to as CEMP-no \citep[and references therein]{BeCh2005,masseron10}. 

The most promising scenario for the formation of these CEMP-s stars states that during their main-sequence phase, they were subject to strong accretion episodes from winds emitted by a more evolved companion. As shown by many authors  \citep{FuIkIb2000,stancliffe08,masseron10,masseron12}, the observed patterns of CEMP-s stars are well accounted for by accretion from an AGB star. The final results depend on several parameters: the composition of the wind, the initial mass of the AGB, the distance of the two stars at the time of accretion, and the age of the system when the accretion occurred.

The situation for other subclasses of CEMP stars is less clear than that of the CEMP-s. The theories include accretion from type II supernova (SNII) transfer, accretion-induced collapse of white dwarfs \citep{jonsell06}, failed SN \citep{umeda03}, and non-standard AGB \citep{suda04}. A statistical analysis of radial velocity variations \citep{lucatello05} indeed shows  that most CEMP-s and CEMP-rs are binaries.

As discussed further on in this paper, the accretion of heavy matter on to a star leads to an inversion of the mean molecular weight, which may trigger thermohaline convection. The resulting abundances depend on the strength and extension of this mixing episode. During its main-sequence evolution, the proto-CEMP star may undergo atomic diffusion, which leads to helium settling that builds a stabilising mu-gradient below the convective zone prior to the accretion episode \citep{thompson08}. This original gradient competes with the one induced by accretion, strongly influencing the mixing and, thus, the final observed abundances.

The lithium abundances are very interesting to study in this context. The observations show a large abundance spread among the CEMP stars, larger than for normal EMP stars \citep{masseron12}. Many main-sequence CEMP stars have lithium abundances smaller than that of the Spite Plateau and some of them have abundances that are compatible with the plateau.

\section{Stellar models}\label{input}
\subsection{Input physics}
The models are computed using the Montr\'eal/Montpellier stellar evolution code. Below we give a brief summary of the input physics used in this code. A complete description may be found in papers by \cite{turcotte98}, \cite{richer98} and \cite{richard01}.

Atomic diffusion is taken into account in all our models for H, $^3$He, $^4$He, $^6$Li, $^7$Li, $^9$Be, $^{10}$B, $^{11}$B, $^{12}$C, $^{13}$C, N, O, Ne, Na, Mg, Al, Si, P, S, Cl, Ar, K, Ca, Ti, Cr, Mn, Fe, and Ni. Here, we compute the diffusion velocities in a more rigorous way than in previous computations of CEMP-s models. We extensively solve the \cite{burgers69} equations, whereas previous works only used the trace element approximation. The formulations of \cite{paquette86} are used to determine the collision integrals. A partial ionisation of elements is taken into account \citep{turcotte98,Schlattl2002}. The radiative accelerations are computed following an opacity sampling method and with OPAL monochromatic data \citep{iglesias96,richer98,ViMiRiRi2010}. The considered initial metallicities for the models used in this work range from [Fe/H]$_\mathrm{ini}$=$-2.31$ (metallicity used in previous studies, see Section \ref{sec:intro}) to metallicities as low as [Fe/H]$_\mathrm{ini}=-5.45$. We adopted an initial distribution for heavy elements as given by \cite{asplund09} and normalised for the considered metallicities. We added an $\alpha$-elements enhancement of [$\alpha$/Fe]=+0.3, as expected for population II stars \citep{VdBetal2000}. The initial lithium abundance is that given by \cite{coc17} from the standard Big Bang nucleosynthesis (SBBN) results, that is, A(Li)=2.74. The complete initial chemical mixtures can be found in Table \ref{tab:0}.

\begin{table}
\centering
\caption{\label{tab:0} Initial chemical compositions of the models. Abundances are in mass fraction.}

\begin{tabular}{l|c|c|c|c}
\noalign{\smallskip}\hline\hline\noalign{\smallskip}

\tiny{[Fe/H]}&-2.31&-3.0&-4.00&-5.45\\
\noalign{\smallskip}\hline\noalign{\smallskip}
H & 0.7537 &  0.7538 & 0.7538 & 0.7538 \\
${^{3}\mathrm{He}}$  & $4.36\times10^{-5}$ &  $4.36\times 10^{-5}$ & $4.36\times10^{-5}$ & $4.36\times10^{-5}$ \\
${^{4}\mathrm{He}}$  &  0.2461 & 0.2461 & 0.2461 & 0.2461 \\
${^{7}\mathrm{Li}}$  &  2.96$\times 10^{-9}$ & 2.96$\times 10^{-9}$ & 2.96$\times 10^{-9}$ & 2.96$\times 10^{-9}$ \\
${^{12}\mathrm{C}}$  &  1.06$\times 10^{-5}$ & 2.12$\times 10^{-6}$ & $2.12\times10^{-7}$ & 7.77$\times 10^{-9}$\\
${^{13}\mathrm{C}}$  &  1.18$\times 10^{-7}$ & 2.36$\times 10^{-8}$ & 2.36$\times 10^{-9}$ & 8.66$\times 10^{-11}$  \\
${\mathrm{N}}$       &  3.10$\times 10^{-6}$ & 6.20$\times 10^{-7}$ & 6.20$\times 10^{-8}$ & 2.27$\times 10^{-9}$\\
${\mathrm{O}}$\tablefoottext{a}       &  5.12$\times 10^{-5}$ & 1.02$\times 10^{-5}$ & 1.02$\times 10^{-6}$ & 3.75$\times 10^{-8}$\\
${\mathrm{Ne}}$\tablefoottext{a}      &  1.13$\times 10^{-5}$ & 2.25$\times 10^{-6}$ & 2.25$\times 10^{-7}$ & 8.26$\times 10^{-9}$\\
${\mathrm{Na}}$\tablefoottext{a}      &  2.61$\times 10^{-7}$ & 5.23$\times 10^{-8}$ & 5.23$\times 10^{-9}$ & 1.92$\times 10^{-10}$\\
${\mathrm{Mg}}$\tablefoottext{a}      &  6.35$\times 10^{-6}$ & 1.27$\times 10^{-6}$ & 1.27$\times 10^{-7}$ & 4.66$\times 10^{-9}$\\
${\mathrm{Al}}$\tablefoottext{b}      &  1.25$\times 10^{-7}$ & 2.50$\times 10^{-8}$ & 2.50$\times 10^{-9}$ & 9.18$\times 10^{-11}$\\
${\mathrm{Si}}$\tablefoottext{a}      &  5.95$\times 10^{-6}$ & 1.19$\times 10^{-6}$ & 1.19$\times 10^{-7}$ & 4.36$\times 10^{-9}$\\
${\mathrm{P}}$\tablefoottext{a}       &  5.21$\times 10^{-8}$ & 1.04$\times 10^{-8}$ & 1.04$\times 10^{-9}$ & 3.82$\times 10^{-11}$\\
${\mathrm{S}}$\tablefoottext{a}       &  2.76$\times 10^{-6}$ & 5.52$\times 10^{-7}$ & 5.52$\times 10^{-8}$ & 2.02$\times 10^{-9}$\\
${\mathrm{Cl}}$\tablefoottext{a}      &  7.31$\times 10^{-8}$ & 1.46$\times 10^{-8}$ & 1.46$\times 10^{-9}$ & 5.36$\times 10^{-11}$\\
${\mathrm{Ar}}$\tablefoottext{a}      &  6.58$\times 10^{-7}$ & 1.32$\times 10^{-7}$ & 1.32$\times 10^{-8}$ & 4.82$\times 10^{-10}$\\
${\mathrm{K}}$\tablefoottext{a}       &  1.36$\times 10^{-8}$ & 2.72$\times 10^{-9}$ & 2.72$\times 10^{-10}$ & 9.99$\times 10^{-12}$\\
${\mathrm{Ca}}$\tablefoottext{a}      &  5.67$\times 10^{-7}$ & 1.13$\times 10^{-7}$ & 1.13$\times 10^{-8}$ & 4.16$\times 10^{-10}$\\
${\mathrm{Ti}}$\tablefoottext{a}      &  2.76$\times 10^{-8}$ & 5.52$\times 10^{-9}$ & 5.52$\times 10^{-10}$ & 2.02$\times 10^{-11}$\\
${\mathrm{Cr}}$      &  7.38$\times 10^{-8}$ & 1.48$\times 10^{-8}$ & 1.48$\times 10^{-9}$ & 5.41$\times 10^{-11}$\\
${\mathrm{Mn}}$\tablefoottext{c}      &  3.43$\times 10^{-8}$ & 6.85$\times 10^{-9}$ & 6.85$\times 10^{-10}$ & 2.51$\times 10^{-11}$\\
${\mathrm{Fe}}$      &  5.81$\times 10^{-6}$ & 1.16$\times 10^{-6}$ & 1.16$\times 10^{-7}$ & 4.26$\times 10^{-9}$\\
${\mathrm{Ni}}$      &  3.18$\times 10^{-7}$ & 6.36$\times 10^{-8}$ & 6.36$\times 10^{-9}$ & 2.33$\times 10^{-10}$\\
\noalign{\smallskip}\hline\noalign{\smallskip}
Z${_\mathrm{ini}}$      &  6.0$\times 10^{-5}$ & 1.2$\times 10^{-5}$ & 1.2$\times 10^{-6}$ & 4.4$\times 10^{-8}$\\
\noalign{\smallskip}\hline\noalign{\smallskip}
\end{tabular}
\tablefoot{see \cite{VdBetal2000}: \tablefoottext{a}{[X/Fe]=$+0.3$,} \tablefoottext{b}{[X/Fe]=$-0.3$,} \tablefoottext{c}{[X/Fe]=$-0.15$.}}
\end{table}

The equation of state is that of Eggleton, Faulkner and Flannery, including Coulomb interactions \citep[CEFF,][]{JCD92}, which takes into account the exact chemical mixture during the stellar evolution, as modified by the atomic diffusion and additional transport processes. Most other equations of state may only be computed for fixed or very limited ranges of mixtures of heavy elements. The Rosseland opacity is computed at each mesh point and each time step from the OPAL monochromatic data \citep{iglesias96}. The opacities are complemented at low temperatures ($T<12000$~K) with the Kurucz opacity tables \citep{Kurucz93CD}. The nuclear reactions rates are from \cite{bahcall92}. We used grey atmospheres as surface boundary conditions.
The convection is treated following the  mixing length theory (MLT)\ of \cite{bohm58} with a solar calibrated mixing-length parameter $\alpha_\mathrm{MLT}=1.66$. 
The  boundaries of the convective zones are based on the Schwarzschild criterion. 

\subsection{Turbulent mixing prescription}
\label{sec:def_turb}

The surface abundance evolution is the result of a complex interaction between microscopic and macroscopic transport processes. Atomic diffusion alone produces too large or too small surface abundances, depending on the elements, compared to those deduced from the observations. Several competing transport processes, such as mass loss \citep{ViMiRiRi2010,michaud11,Vick13,MaSt2016}, turbulence \citep{richer00,richard02,stancliffe09,MaSt2016}, rotation \citep{PaTaChFo2003,talon06,eggenberger10,MaSt2017,deal20}, and thermohaline convection \citep{theado09,deal16} are known. These processes may all be at work together in a given stellar situation, leading to complex mixing. Because the aim of this paper is not to study the individual influence of these processes, we decided to adopt the simple adhoc parametrisation developed in \cite{richer00}, which has been used in many studies \citep{richard01,richard05,michaud11}. The associated turbulent diffusion coefficient is expressed as

\begin{equation}\label{eq_turb}
    D_\mathrm{turb}=400~D_\mathrm{He}(T_0)~\left( \frac{\rho(T_0)}{\rho} \right)^3,
\end{equation}

\noindent where $T_0$ is a reference temperature that may be defined as such, or as the temperature at a reference depth \citep{richer00,richard01,michaud11}. Furthermore, $D_\mathrm{He}(T_0)$ is the diffusion coefficient of helium at the reference point, $\rho$ the local density, and $\rho(T_0)$ the density at the reference point. Here, we use the same parametrisation as \cite[][ie. an amplitude equal to 400 and a power law of 3]{richard05} to be able to compare our results with theirs. As explained in the next section, the reference point is parametrised so as to reproduce a lithium plateau in the considered range of effective temperatures. We note that the reference temperature (or reference depth of the mixing) is the only free parameter in this parametrisation. The models are labelled as T[$\log_{10}(T_0/K)$], as, for instance, T5.80. 

\begin{figure}
\center
\includegraphics[scale=0.65]{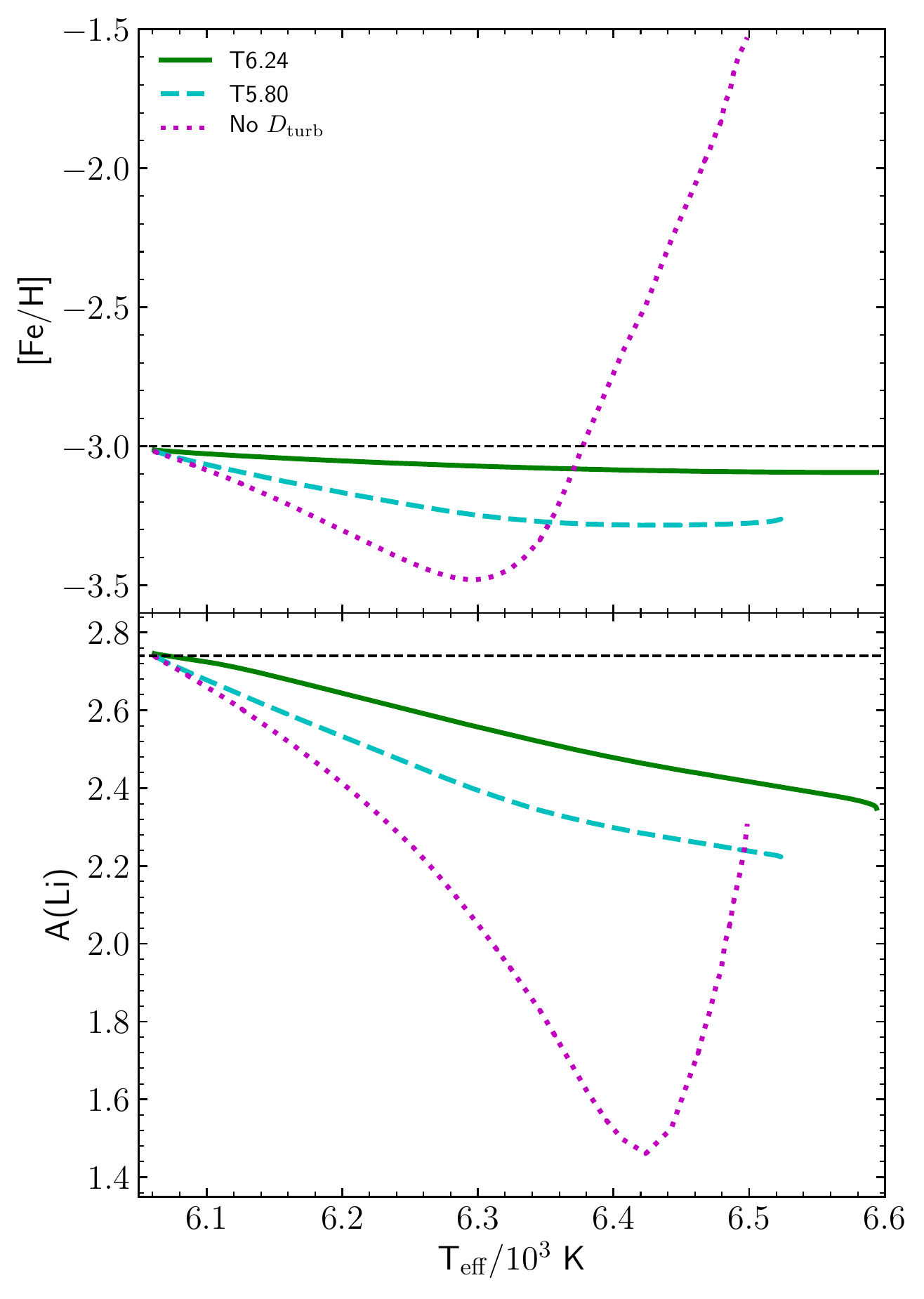}
\caption{Evolution of surface [Fe/H] (top panel) and A(Li) (bottom panel) according to $T_\mathrm{eff}$, for three models with a mass of 0.78\,$M_\odot$ and [Fe/H]$_\mathrm{ini}=-3.0$ on the main sequence. No accretion was taken into account in these models. The magenta dotted line, the blue dashed line and the green line correspond, respectively, to the models with no other mixing than standard convection, with T5.80 turbulence and with T6.24 turbulence. The horizontal black dashed line represents the initial abundances of the models. }
\label{fig:1bis}
\end{figure}

\subsection{Lithium abundances induced by turbulent mixing}

The evolution on the main sequence of surface [Fe/H] and lithium abundance versus $T_\mathrm{eff}$ are presented in Fig. \ref{fig:1bis}, for different prescriptions of turbulence below the convective zone. Atomic diffusion is fully effective in the model in which no other mixing than standard convection is added (magenta dotted lines) and is dominated by the gravitational settling contribution, for both iron and lithium, in the first part of the main-sequence evolution. Around $T_\mathrm{eff}=6300$\,K, the surface convective zone is thin enough that radiative acceleration becomes the dominant contribution for iron but not for lithium, while later when $T_\mathrm{eff}$ rises above $6400$\,K, radiative acceleration becomes the dominant contribution for lithium \citep[see also][]{richard02}.
In the models computed with the turbulence mixing T5.80 and T6.24 (respectively: blue dashed lines and green lines, as described in Section \ref{sec:def_turb}), there are no surface element over-abundances compared to the initial one -- contrary to the models without turbulence. Turbulent mixing prevents any strong element accumulation. The surface abundance of iron is indeed always smaller than the initial one, by less than $0.3$~dex for the T5.80 model and $0.1$~dex for the T6.24 model.

In Fig. 2, we present the behaviour of the lithium surface abundance with $T_\mathrm{eff}$, at $12.5$~Gyr for models without accretion with different masses and [Fe/H]$_\mathrm{ini}$ (different symbols) and with different prescriptions for turbulence below the convective zone. The magenta symbols correspond to models in which no other mixing than standard convection is introduced. In this case, lithium is only subject to atomic diffusion below the convective zone. The lithium abundance decreases with increasing $T_\mathrm{eff}$, due to the dominant contribution of gravitational settling, up to $T_\mathrm{eff}$=6400~K. Above this effective temperature, the lithium abundance increases due to radiative accelerations, which become large enough to overcome gravity for the two lower metallicities. The blue and green symbols correspond respectively to models with T5.80 and T6.25 turbulence below the convective zone. In this range of turbulence prescriptions, the effect of turbulence is twofold. Firstly, it reduces the lithium variations induced by atomic diffusion for hotter stars. We can see in the figure that this effect happens for the two cases, and that it is larger when the turbulence is stronger, as expected. Above 6000~K, the lithium depletion in the T6.24 model is nearly horizontal as a function of $T_\mathrm{eff}$, like a plateau. The second effect of turbulence is that it transports lithium down to its nuclear burning layers in cooler stars. This effect is quite different in the two cases. Lithium is not destroyed for T5.80 models, that is, for the weakest turbulence, whereas it is strongly destroyed in the T6.24 models.

These models show that the reference temperature for the turbulent diffusion coefficient should be between $\log(T_0)=5.8$ and $\log(T_0)=6.24$ to be consistent with a lithium plateau at [Fe/H]$_\mathrm{ini}=-2.31$ down to [Fe/H]$_\mathrm{ini}=-5.45$, which is in agreement with the results of \cite{richard05} for Spite plateau stars. We also show here that the efficiency of the mixing does not depend on the considered initial chemical mixture (\citealt{grevesse93} or \citealt{asplund09}) nor on the metallicity in the range of our study.

\begin{figure}
\center
\includegraphics[scale=0.58]{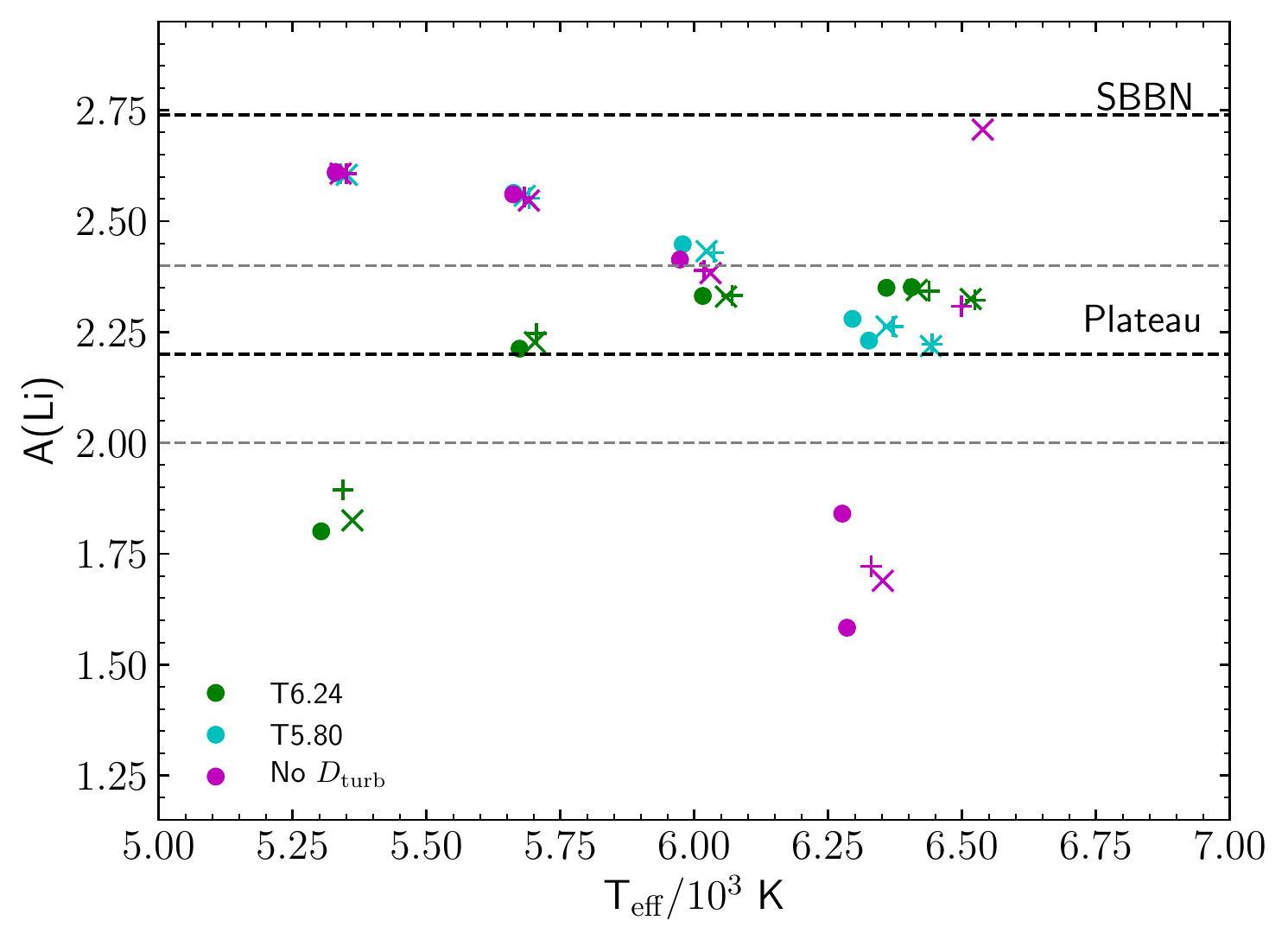}
\caption{Predicted lithium surface abundance according to $T_\mathrm{eff}$, at 12.5\,Gyr with different $D_\mathrm{turb}$ parametrisations, for models without accretion with masses between 0.6 and 0.8\,$M_\odot$ and [Fe/H]$_\mathrm{ini}=-2.31$ (circles), $-3.0$ (+), and $-5.45$ (crosses).}
\label{fig:1}
\end{figure}

\section{Accretion of AGB winds on main-sequence EMP stars }\label{accr}

\subsection{Method}

In the following, we compute the effects of the accretion of AGB winds on main-sequence extremely metal-poor stars, leading to the formation of CEMP-s stars. The proto-CEMP-s models are evolved up to the age at which the more massive companion is going through the AGB phases and ejects matter. The wind matter accreted on to the star is considered as fully mixed inside the convective zone on a time scale smaller than the evolution time step. This accretion is modelled by changing the chemical composition of the convective zone at this age. Then the evolution goes on with the new composition up to the end of the main sequence. The chosen accreted masses are $3.8\times10^{-3}$, $3.8\times10^{-2}$, $3.8\times10^{-1}$, $5.0\times10^{-1}$, $6.0\times10^{-1}$, $8.0\times10^{-1}$, $1.0$, $2.0$, $3.0$, $3.8$, $8,$ and $38$~$M_{\jupiter}$. These modify the total stellar mass by less than about $0.5$\%, which is small enough to be neglected in this type of model, and can be up to about 5\% when 38~$M_{\jupiter}$ are accreted in the case of a companion of 1~$M_\odot$.

\subsection{Chemical composition of the accreted wind}

\begin{table*}[ht]
\centering
\caption{\label{tab:1} Chemical composition of the AGB winds considered in this study, according to the stellar masses and metallicities. The wind abundances are given in mass fraction.}

\begin{tabular}{l|ccc|ccc}
\noalign{\smallskip}\hline\hline\noalign{\smallskip}
\multicolumn{7}{c}{AGB models}\\
\noalign{\smallskip}\hline\noalign{\smallskip}
[Fe/H]$_\mathrm{ini}$ (dex) &\multicolumn{3}{c|}{-2.31\tablefoottext{a}}&\multicolumn{3}{c}{-3.00\tablefoottext{b}}\\
\noalign{\smallskip}\hline\noalign{\smallskip}
Mass/$M_\odot$ & 1.0 & 2.0 & 3.0 & 1.0 & 2.0 & 3.0  \\
\noalign{\smallskip}\hline\hline\noalign{\smallskip}
&\multicolumn{6}{c}{Wind}\\
\noalign{\smallskip}\hline\noalign{\smallskip}
$Z_\mathrm{wind}$ & $4.6\times 10^{-3}$ & $1.5\times 10^{-2}$ & $3.6\times 10^{-3}$ & $2.8\times 10^{-3}$ & $1.4\times 10^{-2}$ & $2.5\times 10^{-2}$ \\
\noalign{\smallskip}\hline\noalign{\smallskip}
H                               & $0.719$& $0.679$ &$0.712$& $0.730$ & $0.678$ & $0.619$  \\
${^{3}\mathrm{He}}$  & $4.35\times 10^{-4}$ & $1.68\times 10^{-4}$ & $4.30\times 10^{-4}$ & $2.28\times 10^{-4}$ & $2.56\times 10^{-6}$&  $3.90\times 10^{-8}$   \\
${^{4}\mathrm{He}}$            & $0.276$& $0.306$ & $0.284$ & $0.267$& $0.308$ & $0.356$    \\
${^{7}\mathrm{Li}}$ & $1.04\times 10^{-11}$ & $1.56\times 10^{-11}$ & $7.79\times 10^{-11}$ &  $2.44\times 10^{-9}$  & $2.01\times 10^{-10}$ & $6.44\times 10^{-11}$    \\
${^{12}\mathrm{C}}$  & $2.76\times 10^{-3}$ & $1.62\times 10^{-2}$ & $5.76\times 10^{-3}$ & $2.01\times 10^{-3}$ & $4.71\times 10^{-4}$ & $8.8\times 10^{-4}$     \\
${^{13}\mathrm{C}}$  & $2.25\times 10^{-6}$ & $7.27\times 10^{-7}$ & $5.48\times 10^{-4}$  & $1.07\times 10^{-4}$  & $1.08\times 10^{-4}$ & $2.04\times 10^{-4}$      \\
${\mathrm{N}}$       & $3.33\times 10^{-5}$ & $2.60\times 10^{-5}$ & $3.69\times 10^{-4}$ & $5.94\times 10^{-5}$ & $1.26\times 10^{-2}$ & $2.23\times 10^{-2}$     \\
${\mathrm{O}}$       & $1.90\times 10^{-4}$ & $4.78\times 10^{-4}$ & $2.48\times 10^{-4}$ & $2.60\times 10^{-4}$ & $1.98\times 10^{-4}$ & $3.48\times 10^{-4}$     \\
${\mathrm{Ne}}$      & $1.02\times 10^{-5}$ & $1.60\times 10^{-5}$ & $1.21\times 10^{-5}$ & $8.40\times 10^{-6}$ & $4.59\times 10^{-5}$ & $2.58\times 10^{-4}$     \\
${\mathrm{Na}}$      & $7.89\times 10^{-7}$ & $1.65\times 10^{-5}$& $2.92\times 10^{-6}$ & $1.58\times 10^{-7}$ & $3.34\times 10^{-4}$ & $6.74\times 10^{-4}$     \\
${\mathrm{Mg}}$      & $2.73\times 10^{-6}$ & $8.56\times 10^{-6}$ & $3.87\times 10^{-6}$ & $1.27\times 10^{-6}$ & $1.43\times 10^{-6}$ & $1.69\times 10^{-6}$    \\
${\mathrm{Al}}$      & $2.92\times 10^{-7}$ & $4.60\times 10^{-7}$ & $8.09\times 10^{-7}$ & $31.6\times 10^{-7}$ & $1.73\times 10^{-6}$ & $1.43\times 10^{-5}$      \\
${\mathrm{Si}}$      & $3.28\times 10^{-6}$ & $3.35\times 10^{-6}$ & $3.65\times 10^{-6}$ & $1.15\times 10^{-6}$ & $1.98\times 10^{-6}$ & $6.30\times 10^{-6}$     \\
${\mathrm{P}}$       & $4.22\times 10^{-7}$ & $4.79\times 10^{-7}$ & $6.00\times 10^{-7}$ & $1.26\times 10^{-6}$ & $7.54\times 10^{-7}$ & $5.44\times 10^{-6}$     \\
${\mathrm{S}}$       & $1.99\times 10^{-6}$ & $1.96\times 10^{-6}$ & $2.02\times 10^{-6}$ & $8.43\times 10^{-6}$ & $6.57\times 10^{-7}$ & $1.70\times 10^{-6}$    \\
${\mathrm{Fe}}$      & $5.83\times 10^{-6}$ & $5.60\times 10^{-6}$ & $5.77\times 10^{-6}$ & $1.39\times 10^{-6}$ & $1.35\times 10^{-6}$ & $1.31\times 10^{-6}$     \\
${\mathrm{Ni}}$      & $2.47\times 10^{-7}$ & $2.33\times 10^{-7}$ & $2.44\times 10^{-7}$ & $2.00\times 10^{-8}$ & $1.90\times 10^{-8}$ & $1.80\times 10^{-8}$     \\
\noalign{\smallskip}\hline\noalign{\smallskip}
\noalign{\smallskip}\hline\noalign{\smallskip}
[Fe/H]$_\mathrm{ini}$ (dex)& \multicolumn{3}{c|}{-4.00\tablefoottext{b}}&\multicolumn{3}{c}{-5.45\tablefoottext{b}}\\
\noalign{\smallskip}\hline\noalign{\smallskip}
Mass/$M_\odot$ & 1.0 & 2.0 & 3.0 & 1.0 & 2.0 & 3.0 \\
\noalign{\smallskip}\hline\hline\noalign{\smallskip}
&\multicolumn{6}{c}{Wind}\\
\noalign{\smallskip}\hline\noalign{\smallskip}
$Z_\mathrm{wind}$ & $7.8\times 10^{-3}$ & $5.0\times 10^{-3}$ & Not available \tablefoottext{b} & $2.8\times 10^{-3}$ & $5.0\times 10^{-3}$ &$1.7\times 10^{-2}$    \\
\noalign{\smallskip}\hline\noalign{\smallskip}
H                               & $0.724$ & $0.695$ & - & $0.730$ & $0.651$ & $0.603$  \\
${^{3}\mathrm{He}}$  & $2.0\times 10^{-4}$ & $2.74\times 10^{-6}$ & - & $1.68\times 10^{-4}$ & $1.71\times 10^{-7}$ &$1.89\times 10^{-8}$   \\
${^{4}\mathrm{He}}$            & $0.268$ & $0.300$ & - & $0.267$ & $0.344$ &$0.380$    \\
${^{7}\mathrm{Li}}$ & $1.73\times 10^{-8}$& $4.43\times 10^{-10}$ & - & $1.00\times 10^{-8}$ & $3.91\times 10^{-10}$ &$5.03\times 10^{-11}$    \\
${^{12}\mathrm{C}}$  & $4.88\times 10^{-3}$ & $1.77\times 10^{-4}$ & - & $1.84\times 10^{-3}$ & $4.96\times 10^{-4}$ & $6.35\times 10^{-4}$    \\
${^{13}\mathrm{C}}$  & $3.69\times 10^{-4}$ & $4.30\times 10^{-5}$ & - & $1.42\times 10^{-4}$ & $1.14\times 10^{-4}$ & $1.49\times 10^{-4}$    \\
${\mathrm{N}}$       & $2.06\times 10^{-4}$ & $4.41\times 10^{-3}$ & - & $8.07\times 10^{-5}$ & $1.34\times 10^{-2}$ & $1.58\times 10^{-2}$    \\
${\mathrm{O}}$       & $1.73\times 10^{-3}$ & $9.31\times 10^{-5}$ & - & $1.04\times 10^{-4}$ & $2.06\times 10^{-4}$ & $2.23\times 10^{-4}$    \\
${\mathrm{Ne}}$      & $2.06\times 10^{-5}$ & $7.59\times 10^{-5}$ & - & $3.80\times 10^{-6}$ & $5.82\times 10^{-5}$ & $2.17\times 10^{-4}$    \\
${\mathrm{Na}}$      & $4.95\times 10^{-6}$ & $4.97\times 10^{-5}$ & - & $1.61\times 10^{-7}$ & $2.50\times 10^{-4}$ & $2.61\times 10^{-4}$    \\
${\mathrm{Mg}}$      & $3.16\times 10^{-7}$ & $2.26\times 10^{-7}$ & - & $5.94\times 10^{-9}$ & $3.10\times 10^{-7}$ & $6.03\times 10^{-7}$    \\
${\mathrm{Al}}$      & $2.97\times 10^{-8}$ & $4.37\times 10^{-7}$ & - & $1.61\times 10^{-10}$ & $2.63\times 10^{-6}$ & $9.79\times 10^{-6}$    \\
${\mathrm{Si}}$      & $1.66\times 10^{-7}$ & $2.30\times 10^{-7}$ & - & $4.32\times 10^{-9}$ & $1.02\times 10^{-6}$ & $2.43\times 10^{-6}$    \\
${\mathrm{P}}$       & $4.88\times 10^{-8}$ & $8.55\times 10^{-8}$ & - & $8.49\times 10^{-11}$ & $8.33\times 10^{-7}$ & $2.47\times 10^{-6}$    \\
${\mathrm{S}}$       & $9.00\times 10^{-8}$ & $7.05\times 10^{-8}$ & - & $1.84\times 10^{-9}$ & $1.82\times 10^{-7}$ & $5.96\times 10^{-7}$    \\
${\mathrm{Fe}}$      & $1.38\times 10^{-7}$ & $1.38\times 10^{-8}$ & - & $5.00\times 10^{-9}$ & $4.91\times 10^{-9}$ & $4.82\times 10^{-9}$    \\
${\mathrm{Ni}}$      & $1.94\times 10^{-9}$ & $1.95\times 10^{-9}$ & - & $7.07\times 10^{-11}$ & $6.92\times 10^{-11}$ & $6.80\times 10^{-11}$ \\
\noalign{\smallskip}\hline
\end{tabular}

\tablefoot{\tablefoottext{a}{\cite{stancliffe08}},\tablefoottext{b}{\cite{campbell08}} }
\end{table*}

For the chemical composition of the accreted matter, we have selected that given in the literature for AGB winds, according to the metallicity range of the CEMPS-s stars by, namely, \cite{stancliffe08} for [Fe/H]$_\mathrm{ini}$=$-2.31$ and \cite{campbell08} for the lower metallicities of [Fe/H]$_\mathrm{ini}$=$-3.00$, $-4.00$ and [Fe/H]$_\mathrm{ini}$=$-5.45$. We only considered the elements  taken into account in the Montr\'eal/Montpellier code. The considered chemical compositions and ejection ages are presented in Table \ref{tab:1} for the different masses of AGB companions used. We note that these wind compositions do not take into account the possible increase of lithium inside the wind due to the thermohaline convection that may occur during the AGB phase as shown by \cite{stancliffe10}. The impact of neglecting this aspect is discussed in Section \ref{LiAGB}. 

\subsection{Age at the time of accretion}

Another important aspect is the age at which the winds are ejected, that is, when they are accreted by the  proto-CEMP-s. \cite{stancliffe08} provided ages at which the ejection of AGB winds end. We considered these ages as the time when all the accreted matter reaches the surface convective zone of the accreting star. This assumption is valid as long as the time step of evolution of the accreting star (in our models, between 10 and 150 Myr on the main sequence) is on the same order as or larger than the typical time of the AGB phase of the companion, namely, between $10$ and $17$ Myr at 0.9~$M_\odot$, decreasing with increasing mass, as given by \cite{bressan86} and \cite{bossini15} or between $1.3$ and $0.32$~Myr, according to \cite{dellagli19}. From the point of view of the evolution of the accreting star model, this can be considered as instantaneous accretion over a time step. 

\subsection{Thermohaline convection}

Thermohaline (fingering) convection occurs when the thermal gradient is stable and the mean molecular weight gradient unstable. If the thermal stabilising effect is smaller than that of the $\mu$ destabilising, this leads to dynamical convection (Ledoux criterion). In the reverse case, the medium should be stable in the first approximation but it is not; this is due to the fact that the thermal diffusivity is much larger than the particle diffusivity. When a blob of matter falls towards the centre of the star, the heat goes in more quickly than the particles escape and the blob keeps on sinking. This effect generates a mixing of chemicals in that region. Such an instability may occur during the main-sequence phase in the case of the accumulation of local elements by atomic diffusion \citep{theado09,deal16,hui-Bon-Hoa18} or in the case of accretion \citep[][and reference therein]{vauclair04,stancliffe07,theado10,garaud11,deal13,deal15,wachlin17}. In the low-mass models of this study, atomic diffusion does not lead to any unstable mean molecular weight gradient. On the other hand, the accretion of heavy matter is able to produce such a gradient and trigger thermohaline convection.

We implemented the 1D prescription of \cite{brown13} in the Montr\'eal/Montpellier code in the same way as it was applied in the Toulouse Geneva Evolution code \citep{zemskova14,deal15,deal16}. This prescription has been shown to be in better agreement with 3D than the previous simulations. In particular, the study of \cite{kippenhahn80} overestimates the mixing efficiency in some configurations \citep{prat15}. Here, we do not include the possible effects of shears \citep{garaud19} or magnetic fields \citep{harrington19}.

\begin{figure}[ht]
\center
\includegraphics[scale=0.68]{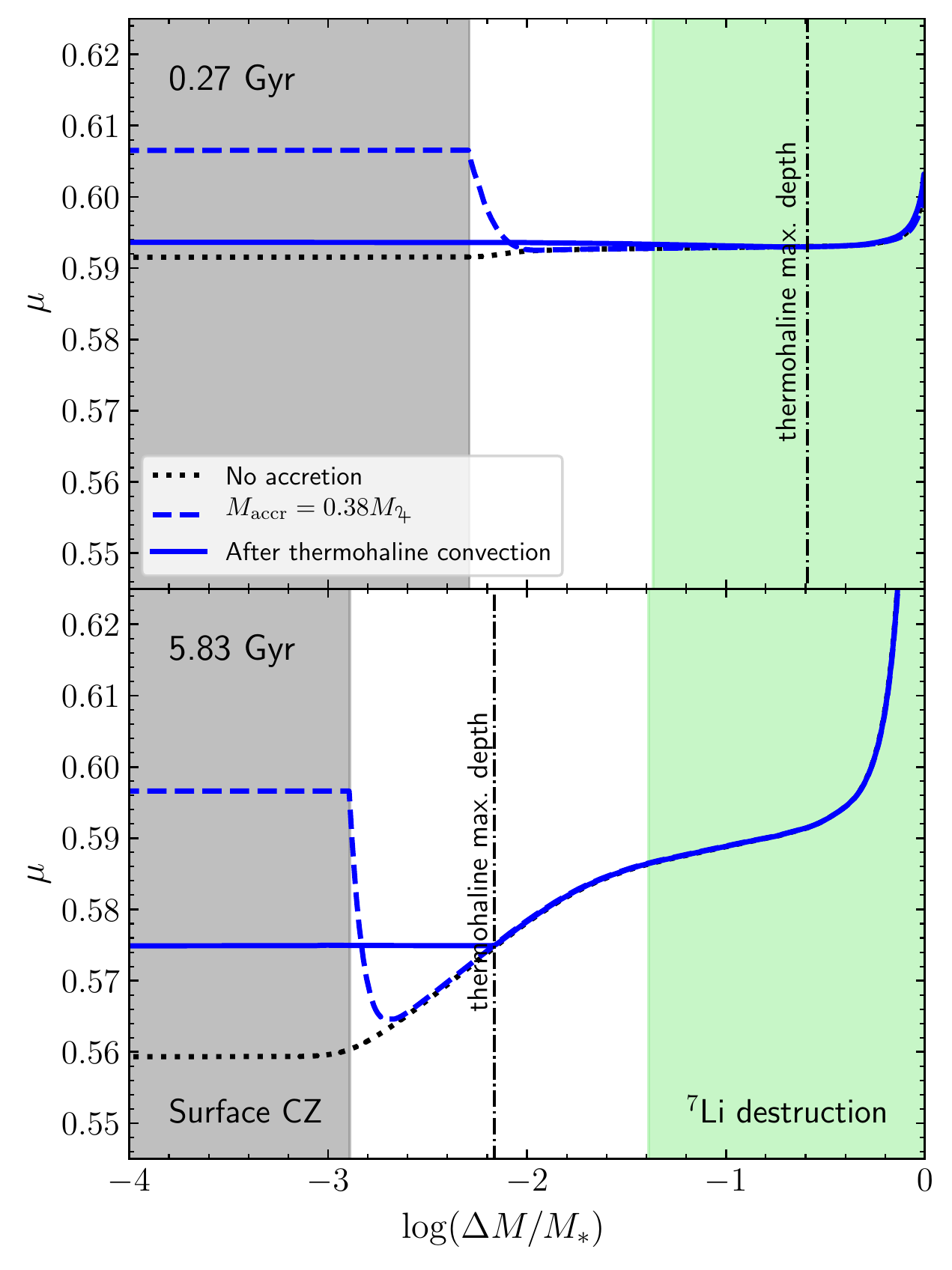}
\caption{$\mu$-profiles inside a 0.8~$M_\odot$ model at two ages. The black dotted lines are the $\mu$ profiles without accretion, the blue dashed lines represent the situation just after the accretion phase, and the blue solid lines are after the thermohaline convection mixing. The grey area represents the surface convective zones and the green one the region of lithium nuclear destruction. The vertical dashed-dotted lines represent the maximum depth of the thermohaline mixing.}
\label{fig:2}
\end{figure}

\begin{table*}[ht]
\centering
\caption{\label{tab:2} Characteristics of the CEMP models with T6.24 turbulent mixing.}

\begin{tabular}{l|ccc|ccc|ccc}
\noalign{\smallskip}\hline\hline\noalign{\smallskip}
&\multicolumn{9}{c}{[Fe/H]$_\mathrm{ini}$=-2.31 | $Z_\mathrm{ini}=6.0\times10^{-5}$ | X($^{12}$C)$_\mathrm{ini}=1.1\times10^{-5}$}\\
\noalign{\smallskip}\hline\noalign{\smallskip}
AGB mass / $M_\odot$ &\multicolumn{3}{c|}{1.0}&\multicolumn{3}{c|}{2.0}&\multicolumn{3}{c}{3.0}\\
Accretion age / Gyr&\multicolumn{3}{c|}{5.83}&\multicolumn{3}{c|}{0.748}&\multicolumn{3}{c}{0.270}\\
\noalign{\smallskip}\hline\noalign{\smallskip}
CEMP mass / $M_\odot$ & 0.70 & 0.75 & 0.78 & 0.70 & 0.75 & 0.78 & 0.70 & 0.75 & 0.78\\

$M_\mathrm{CZ}$/$M_{\jupiter}$\tablefoottext{a}  & 6.67 & ~1.89~ & 0.62~ & ~11.0~ & ~4.64~ & ~2.45~~ & ~11.74~ & ~4.95~ & ~2.73~~\\
$M_\mathrm{mix}$/$M_{\jupiter}$\tablefoottext{b}  & 6.67 & ~6.68~ & 6.25~ & ~~11.0 ~ & ~6.27~ & ~5.90~~ & ~~11.74 ~ & ~6.25~ & ~5.86~~\\

\noalign{\smallskip}\hline\hline\noalign{\smallskip}
&\multicolumn{9}{c}{[Fe/H]$_\mathrm{ini}$=-3.00 | $Z_\mathrm{ini}=1.2\times10^{-5}$ | X($^{12}$C)$_\mathrm{ini}=2.1\times10^{-6}$}\\
\noalign{\smallskip}\hline\noalign{\smallskip}
AGB mass / $M_\odot$&\multicolumn{3}{c|}{1.0}&\multicolumn{3}{c|}{2.0}&\multicolumn{3}{c}{3.0}\\
Accretion age / Gyr&\multicolumn{3}{c|}{5.83}&\multicolumn{3}{c|}{0.748}&\multicolumn{3}{c}{0.270}\\
\noalign{\smallskip}\hline\noalign{\smallskip}
CEMP mass / $M_\odot$ & 0.70 & 0.75 & 0.78 & 0.70 & 0.75 & 0.78 & 0.70 & 0.75 & 0.78\\
$M_\mathrm{CZ}$/$M_{\jupiter}$\tablefoottext{a}   & 6.11 & 1.59 & 0.48 & 10.20 & 4.17 & 2.21 & 10.73 & 4.42 & 2.32\\
$M_\mathrm{mix}$/$M_{\jupiter}$\tablefoottext{b}  & 6.96 & ~6.60~ & 6.42 & ~~10.20 ~ & ~6.29~ & ~6.28~ & ~~10.73 ~ & ~6.26~ & ~6.22~~\\

\noalign{\smallskip}\hline\hline\noalign{\smallskip}
&\multicolumn{9}{c}{[Fe/H]$_\mathrm{ini}$=-4.00 | $Z_\mathrm{ini}=1.2\times10^{-6}$ | X($^{12}$C)$_\mathrm{ini}=2.1\times10^{-7}$}\\
\noalign{\smallskip}\hline\noalign{\smallskip}
AGB mass / $M_\odot$&\multicolumn{3}{c|}{1.0}&\multicolumn{3}{c|}{2.0}&\multicolumn{3}{c}{-}\\
Accretion age / Gyr&\multicolumn{3}{c|}{5.83}&\multicolumn{3}{c|}{0.748}&\multicolumn{3}{c}{-}\\
\noalign{\smallskip}\hline\noalign{\smallskip}
CEMP mass / $M_\odot$ & 0.70 & 0.75 & 0.78 & 0.70 & 0.75 & 0.78 & - & - & - \\
$M_\mathrm{CZ}$/$M_{\jupiter}$\tablefoottext{a}  & 6.05 & 1.55 & 0.45 & 9.96 & 4.15 & 2.14 & - & - & -\\
$M_\mathrm{mix}$/$M_{\jupiter}$\tablefoottext{b}  & 6.99 & ~6.54~ & 6.32 & ~9.96~ & ~6.24~ & ~~6.07~~ & - & - & -\\

\noalign{\smallskip}\hline\hline\noalign{\smallskip}
&\multicolumn{9}{c}{[Fe/H]$_\mathrm{ini}$=-5.45 | $Z_\mathrm{ini}=4.4\times10^{-8}$ | X($^{12}$C)$_\mathrm{ini}=7.8\times10^{-9}$}\\
\noalign{\smallskip}\hline\noalign{\smallskip}
AGB mass / $M_\odot$&\multicolumn{3}{c|}{1.0}&\multicolumn{3}{c|}{2.0}&\multicolumn{3}{c}{3.0}\\
Accretion age / Gyr&\multicolumn{3}{c|}{5.83}&\multicolumn{3}{c|}{0.748}&\multicolumn{3}{c}{0.270}\\
\noalign{\smallskip}\hline\noalign{\smallskip}
CEMP mass / $M_\odot$ & 0.70 & 0.75 & 0.78 & 0.70 & 0.75 & 0.78 & 0.70 & 0.75 & 0.78\\
$M_\mathrm{CZ}$/$M_{\jupiter}$\tablefoottext{a}   & 5.87 & 1.51 & 0.44 & 9.90 & 4.08 & 2.12 & 10.27 & 4.40 & 2.29\\
$M_\mathrm{mix}$/$M_{\jupiter}$\tablefoottext{b}  & 6.92 & ~6.47~ & 6.31 & ~9.90~ & ~6.35~ & ~6.22~ & ~10.27~ & ~6.22~ & ~6.22~~\\
\noalign{\smallskip}\hline\noalign{\smallskip}
\end{tabular}
\tablefoot{\tablefoottext{a}{At the time of accretion.}\tablefoottext{b}{Masses of the mixed zones defined as the regions where $D_\mathrm{turb}/D_\mathrm{He}\geq 1000$ (see Eq. \ref{eq_turb}). When $D_\mathrm{turb}$ is too small at the bottom of the surface convective zone to satisfy this condition we set} $M_{mix}=M_{CZ}$.
} 
\end{table*}

\section{Results of the accretion of AGB winds on the stellar chemical composition}

We now present the results we obtained on the modifications of the chemical surface abundances of EMP main-sequence stars subject to the wind of AGB companions. Following their evolutionary stages, the AGB is referred to as the primary star of the system and the EMP star (proto-CEMP star) as the secondary. As discussed above, we take into account the chemical modification that is initially induced by the accretion. This depends on the AGB mass, the composition of the wind, the distance between the two stars, and the convective mass of the secondary. At the same time, we introduce the mixing induced by thermohaline convection, which depends on the internal structure of the secondary, and on its age.

\subsection{The formation of CEMP-s stars and lithium abundance}
\label{sec:mu}

In order to assess the impact of thermohaline convection induced by accretion on the surface abundances of EMP stars, we computed models at three initial metallicities, for two masses of the primary (the AGB), three masses of the secondary (the proto-CEMP-s), and for accreted masses between $3.8\times10^{-3}$ and 38\,$M_{\jupiter}$. The characteristics of the models are presented in Table~\ref{tab:2}.

As shown in Tables \ref{tab:1} and \ref{tab:2}, in all cases, the initial total metallicity (Z$_\mathrm{ini}$) of the secondary (the proto-CEMP-s) is between two and seven orders of magnitude lower that the metallicity of the wind (Z$_\mathrm{wind}$). This means that small amounts of matter accreted by the proto-CEMP-s stars are enough to induce unstable mean molecular weight gradients at the bottom of the surface convective zones and lead to thermohaline convection in these regions. The $^{12}$C relative abundance is at least two orders of magnitude larger in the wind than at the surface of the EMP star. The amount of accreted matter needed to create a CEMP star is then quite small. As for lithium, the final abundances strongly depend on the stabilising $\mu$-gradients that the EMP stars had time to develop before the accretion episodes.

For example, for a 1~$M_\odot$ star with [Fe/H]$_\mathrm{ini}$=$-2.31$ ($Z_\mathrm{ini}=6.0\times10^{-5}$), \cite{stancliffe08} predicted an AGB wind ejection at the age of $5.83$~Gyr. Up to that age, the main-sequence proto-CEMP-s star had enough time to build a strong stabilising $\mu$-gradient below the outer convective zone, basically due to helium settling (lower panel of Fig. \ref{fig:2}). After accretion, thermohaline convection may only develop down to the stabilising $\mu$-gradient, that is, $\log(M/M_\ast)\approx -2.2$. This leads to heavy element overabundances in the stellar outer layers, but no extra lithium destruction, because the mixing process did not reach the lithium nuclear destruction layers. The accreting star becomes a CEMP-s star with surface lithium close to the plateau abundance.

For a 3~$M_\odot$ star with the same metallicity, [Fe/H]$_\mathrm{ini}$=$-2.31$, the AGB wind ejection occurs much earlier, at an age of 270 Myr. In this case, the proto-CEMP-s star did not have enough time to build a strong stabilising gradient before the accretion episode. The induced thermohaline convection is not prevented from proceeding deep into the star, so that the accreted matter is mixed down to the lithium destruction region (upper panel of Fig. \ref{fig:2}). The accreting star becomes a CEMP-s star with depleted lithium at the surface. This confirms the results that had previously been obtained \citep[eg.][]{stancliffe08}.

\begin{figure*}
\center
\includegraphics[scale=0.89]{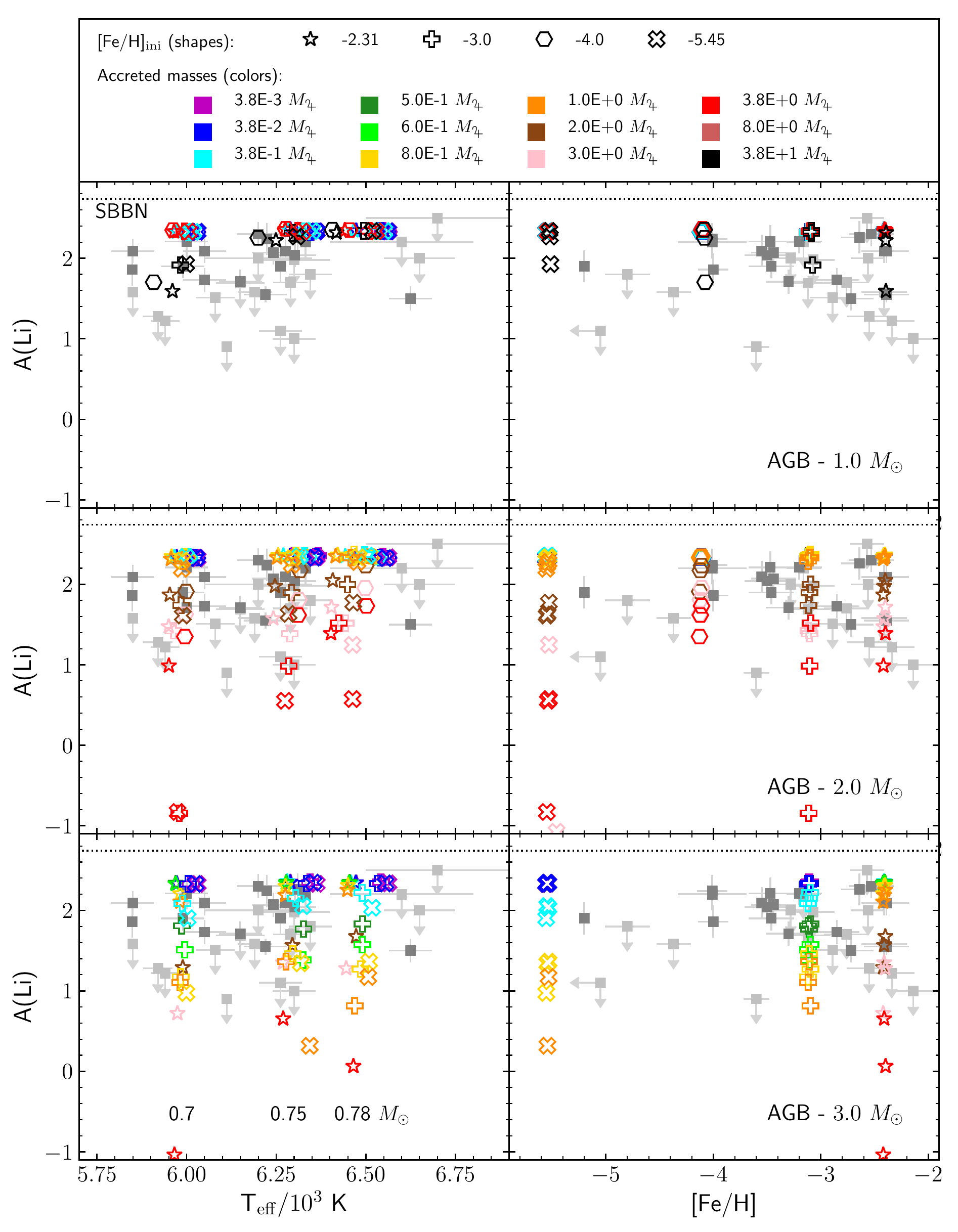}
\caption{Lithium surface abundances obtained in our models at the age of $12.5$\,Gyr. Each model includes an accretion episode of AGB wind at a specific age depending on the AGB mass. The left column presents the Li results according to the stellar effective temperature. The vertical accumulations of symbols correspond to proto-CEMP models of respectively $0.7$, $0.75$ and $0.78$~$M_\odot$. The various metallicities of the models are represented by the shapes of the open symbols and the accreted masses by their various colours. In the right column, the lithium surface abundances are displayed as a function of the [Fe/H] ratio obtained in the same stars at the same age. The grey squares represent the observations (see text for details), light-grey squares are upper limits. The black dotted line is the SBBN lithium abundance that has been used as the initial abundance in our computations.}

\label{fig:3}
\end{figure*}

\begin{figure}
\center
\includegraphics[scale=0.81]{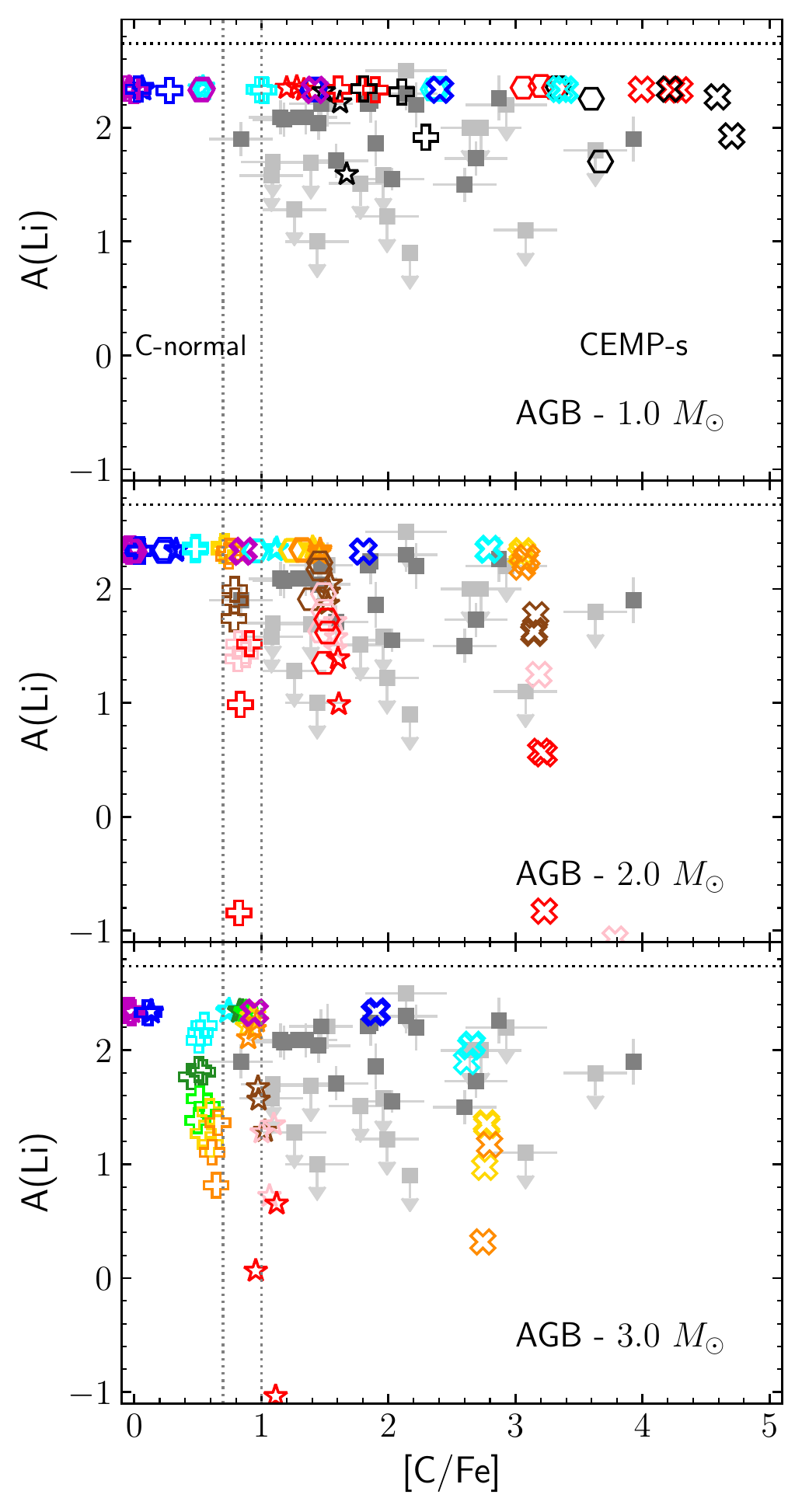}
\caption{Lithium surface abundances obtained in our models at the age of 12.5\,Gyr as a function of [C/Fe]. The vertical dotted lines correspond to [C/Fe]=0.7 and 1.0. The symbols are the same as in Fig. \ref{fig:3}.}
\label{fig:3bis}
\end{figure}

We present in Fig.~\ref{fig:3} the results of the predicted abundances of our models together with the most recent observations of the surface lithium abundances in CEMP-s main-sequence stars and in some unclassified CEMP stars, as gathered in the SAGA data base\footnote{http://sagadatabase.jp/} (\citealt{suda08,suda11,yamada13,suda17} and \citealt{matsuno17}). We notice that around half of observed lithium abundances are, in fact, upper limits, so that the real lithium abundance may still be smaller than the given points.

The surface lithium abundances shown with open symbols in this figure are those predicted by our models at an age of $12.5$~Gyr. Each model has undergone an accretion episode at a time related to the evolution and, thus, to the mass of the AGB companion (top panels: 1.0~$M_\odot$, accretion at $5.83$\,Gyr; middle panels: $2.0$~$M_\odot$, accretion at $0.748$\,Gyr; bottom panels: 3.0~$M_\odot$, accretion at $0.27$\,Gyr). The left and the right panels display, respectively, the results as a function of the effective temperature and metallicity for the three considered masses of the proto-CEMP-s, that is, 0.7, 0.75, and 0.78~$M_\odot$. We can see that an important lithium dispersion appears in the models for AGB masses of 2.0~$M_\odot$ and 3.0~$M_\odot$, which is in accordance with the observations. 

The reason for the appearance of lithium destruction for large-enough AGB masses is indirect. The larger the AGB mass, the more rapid its evolution and the smaller the age of the proto-CEMP-s when the accretion episode occurs. For small ages, the stabilising $\mu-$gradient did not have enough time to settle below the convective zone, so that thermohaline convection was not prevented from proceeding down to the lithium destruction layer. For the 3.0~$M_\odot$ AGB (earliest accretion), with accreted masses between $0.38$ and $1~M_{\jupiter}$, the models predict depletion of lithium down to A(Li) $\approx0.5$. For the 2.0~$M_\odot$ AGB slightly higher accreted masses, between $1$ and $3.8~M_{\jupiter}$, are needed to obtain a similar depletion. For the 1.0~$M_\odot$ AGB, the models predict depletion of lithium for accreted masses of at least $38~M_{\jupiter}$. For larger accreted masses, lithium would be even more depleted because a larger $\mu$-gradient inversion induces a more efficient thermohaline convection.

Figure~\ref{fig:3bis} is similar to Fig.~\ref{fig:3} and presents the lithium surface abundances as a function of [C/Fe]. The carbon-over-iron ratio [C/Fe] increases according to the accreted mass due to the chemical composition of the AGB wind, but its final value also depends on the mixing process below the convective zone, which dilutes the abundances downwards inside the star. We can also see in Fig. \ref{fig:3bis} that the maximum [C/Fe] which is reached decreases for increasing AGB mass. Here again, the reason is related to the age at which the accretion episode occurs. For large AGB masses, thermohaline convection is not prevented from going deep into the star, so that carbon is more diluted. This is also the reason why, for an AGB mass of 3.0~$M_\odot$, lithium destruction appears for [C/Fe] ratios smaller than $0.7$: deep mixing leads both to lithium depletion and carbon dilution. We note that the variation of [C/Fe] for each case depends on the detailed chemical composition of the wind and of the metallicity of the considered stars. On the contrary, the lithium depletion mainly depends on the mean molecular weight of the accreted matter.

The lithium abundance in the wind may vary according to  the metallicity and the mass of the AGB star. However, in the cases where lithium is not destroyed by nuclear reactions, all models lead to lithium abundances that are very close to those of the plateau. The reason is that the mass of the region where the chemical composition is homogenised by the turbulent mixing is similar in all models and much larger than the accreted masses (see $M_\mathrm{mix}$ in Table 3). In this situation, the total mixed zone is much more important than the dynamical convective zone of the proto-CEMP-s and the lithium abundance differences are smoothed out by dilution.

The proto-CEMP stars also accrete s-process elements from the AGB wind so that they become CEMP-s stars. We do not include a detailed study of the evolution of s-process elements in these stars in this paper and we limit our discussion to the evolution of a few heavy elements below.

\subsection{Heavy elements and isotopic carbon ratios}
\begin{figure}
\center
\includegraphics[scale=0.58]{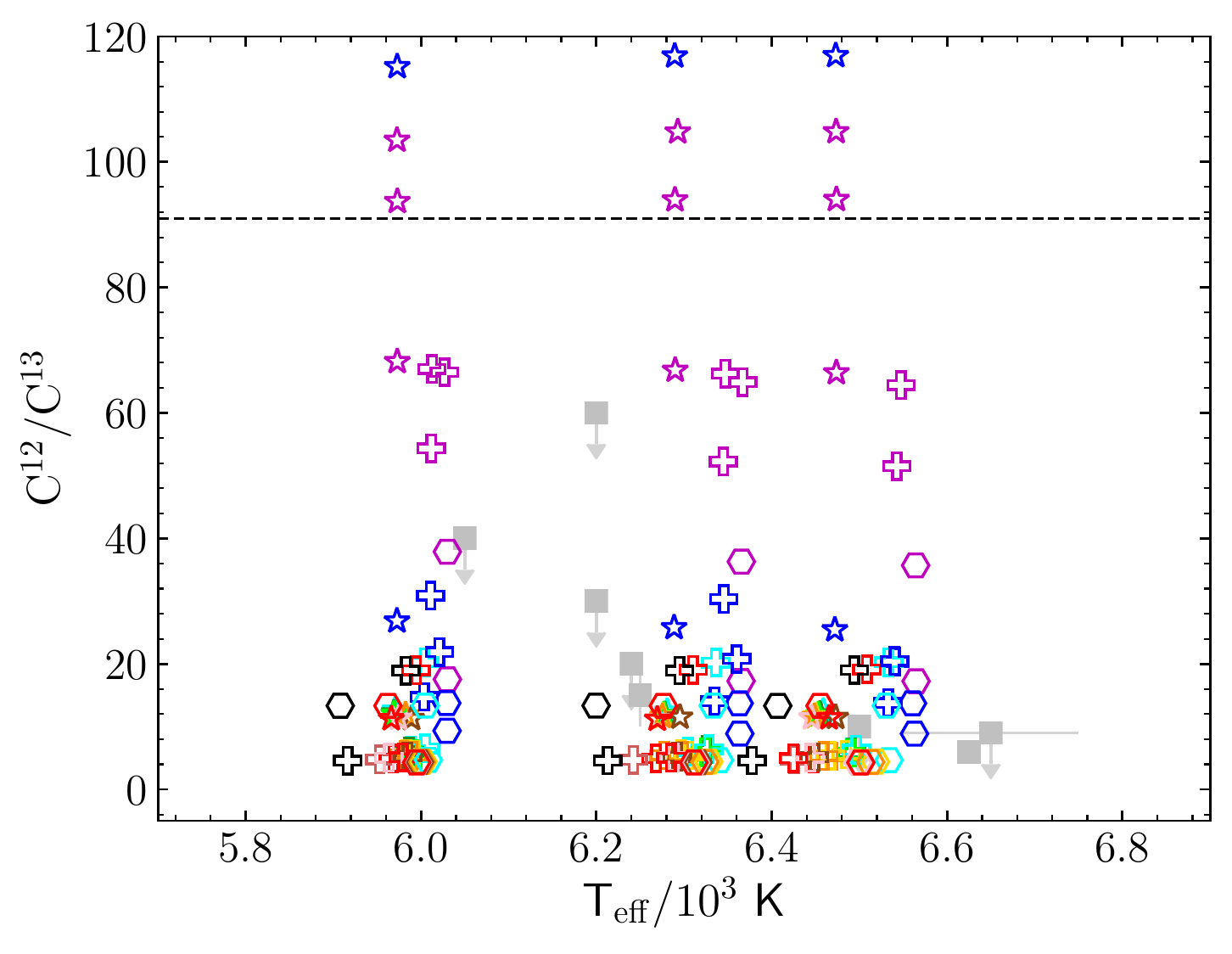}
\caption{$^{12}\mathrm{C}/^{13}\mathrm{C}$ according to the effective temperature. The legend is the same as Fig. \ref{fig:3}. The horizontal dashed line represent the $^{12}C/^{13}C$ value of models without accretion. Only models with [Fe/H]~$>-5.45$ are represented to be consistent with the metallicity of the available observed stars.}
\label{fig:4}
\end{figure}

\begin{figure}
\center
\includegraphics[scale=0.7]{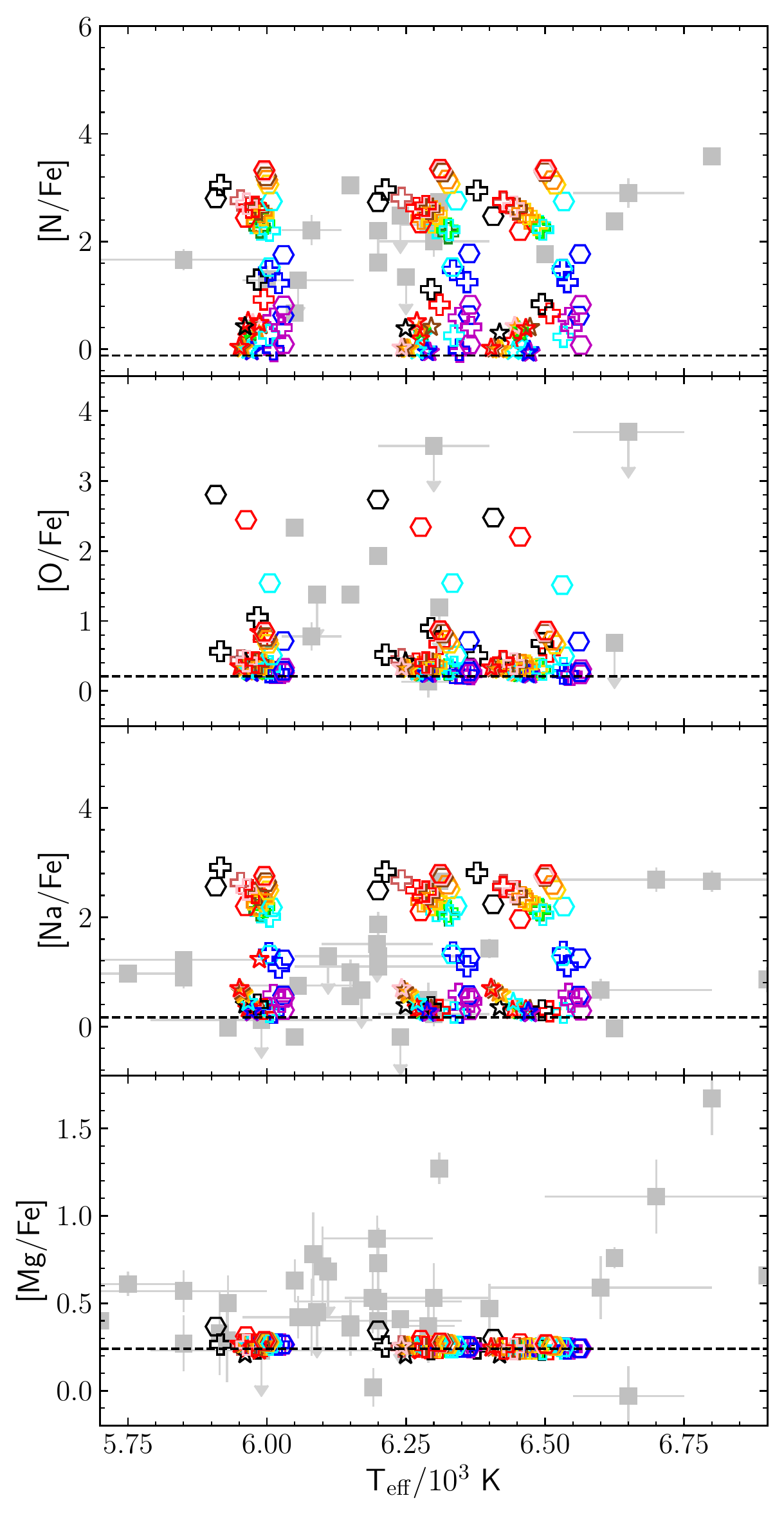}
\caption{Surface abundance ratio [X/Fe] for N, O, Na, and Mg according to effective temperature. The legend is the same as Fig. \ref{fig:3}. The horizontal dashed line represents [X/Fe] from models without accretion. Only models with [Fe/H]~$>-5.45$ are represented so as to be consistent with the metallicity of the available observed stars.}
\label{fig:5}
\end{figure}

In this section, we only consider models with [Fe/H]~$>-5.45$ so as to be consistent with the metallicity of the stars for which the considered chemical elements are observed.

The ratio $^{12}$C/$^{13}$C is smaller in CEMP-s stars than in C-normal stars. Figure \ref{fig:4} presents the comparison between our models and the observations from the SAGA data base (same references as in Fig.~\ref{fig:3}). CEMP-s main-sequence stars have $^{12}$C/$^{13}$C between $0$ and $60$. Models without accretion show larger values of $^{12}$C/$^{13}$C$\approx 91$. On the other hand, the observed ratios are easily reached when accretion is considered in the models. The larger ratios (greater than $10$) are obtained for CEMP-s models formed from the accretion from the 1~$M_\odot$ AGB companion.

We also compare the prediction of our models with the surface abundances of N, O, Na, and Mg obtained from observations (Fig. \ref{fig:5}). The surface abundances [N/Fe] predicted by models with accretion and those deduced from observations are in good agreement, whereas models without accretion are not able to reproduce such large [N/Fe] ratios. Oxygen abundances are available in the same metallicity range and in this case, the models are also able to reproduce the observations, considering the upper limits. Sodium is also well explained by models with the same metallicity range. Our results down to [Fe/H]$=-4.0$ are in agreement with those of previous studies at [Fe/H]$=-2.31$ \citep{stancliffe07,stancliffe08,stancliffe09,stancliffe09bis}.

The models are not able to explain the larger [Mg/Fe] ratios. The discrepancy between CEMP-s models obtained from the accretion of an AGB companion for Mg was also pointed out by \cite{stancliffe09}. This indicates either a possible issue in the predicted chemical abundances in AGB winds or in the accretion/transport scenario. The fact that lithium surface abundances are well reproduced in our models may point to the first explanation because lithium is a powerful probe of the transport of chemical elements in stars. 

\section{Discussion and conclusions}\label{discu}
\subsection{Impact of the input physics on the computed lithium abundances}

\begin{figure}
\center
\includegraphics[scale=0.58]{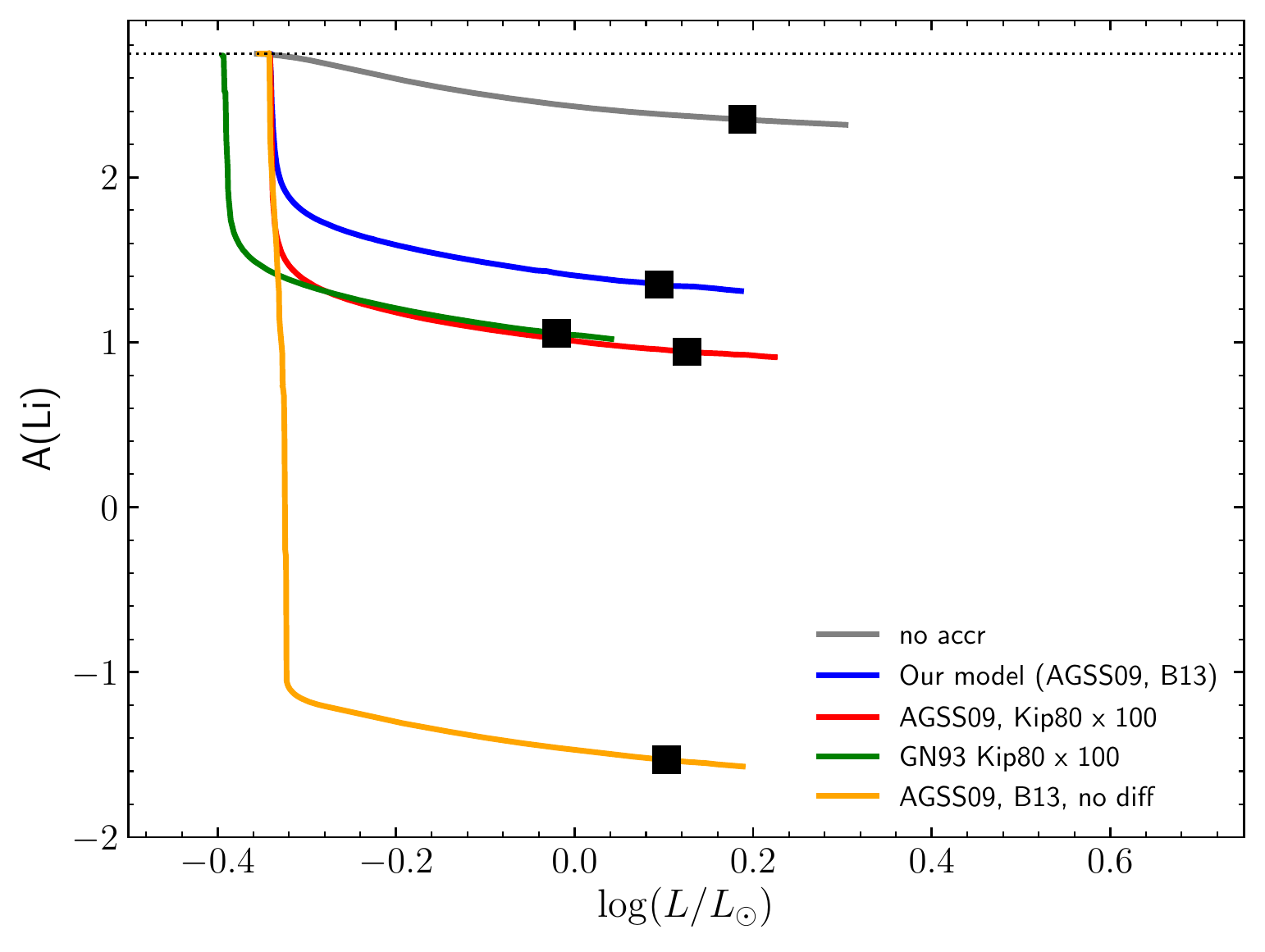}
\caption{Evolution of lithium surface abundance in 0.75\,$M_\odot$ models with [Fe/H]$_\mathrm{ini}=-2.31$ for different input physics. The grey solid curve represents a model without accretion. All the other curves correspond to models of accreting stars with an accreted mass of 3~$M_{\jupiter}$.The blue curve represents a model including the physics used in this study namely AGSS09 for \citealt{asplund09} initial metal mixture, B13 for \citealt{brown13} thermohaline convection prescription, and atomic diffusion (see Section~\ref{input}). The red curve represents a model taking into account the \cite{kippenhahn80} prescription for thermohaline convection multiplied by 100 (Kip80$~\times 100$) and an AGSS09 initial mixture of metals. The green solid curve shows a model with the same prescription for thermohaline convection and an initial mixture of metals following \cite{grevesse93} (GN93). The orange curve represents a model with a similar physics than the blue model but without atomic diffusion (no diff). The black squares stand for the lithium surface abundance at $12.5$~Gyr. The dotted horizontal line indicates the initial lithium abundance.}
\label{fig:physics}
\end{figure}

The results presented in this paper show that the observations of lithium abundances in extremely metal-poor stars may be accounted for in the framework of the accretion scenario. A main-sequence metal-poor star accretes matter from the wind of an AGB companion. This creates an inversion of the mean molecular weight and leads to thermohaline convection. The effect of this convection depends on the accreted mass, but also on the physical state of the accreting star before accretion. If this star had time to build a stabilising helium gradient under its convective zone, the thermohaline mixing is less efficient. Our results differ from the previously published ones for several reasons, including the input physics, the range of metallicities of the studied stars, and the range of accreted masses.

Concerning the input physics, basic improvements may be found in the computation of atomic diffusion and that of thermohaline mixing. In our code, the treatment of atomic diffusion is done by solving the full set of Burgers equations, whereas previous studies remained in the framework of the trace element approximation. To show the importance of atomic diffusion in the final results, we present in Fig.~\ref{fig:physics} computations of lithium abundance variations during the evolution of a $0.75$~$M_\odot$, [Fe/H]$_\mathrm{ini} = -2.31$, in which we have suppressed atomic diffusion (orange curve, compared to the blue one done with the correct physical input). There is a difference of about 3~dex for the lithium abundance obtained at $12.5$~Gyr.

For thermohaline convection, we used the prescription of \cite{brown13}, which is based on 3D simulations. Previous studies used either the \cite{kippenhahn80} approximate diffusion coefficient, or that same coefficient multiplied by 100 as proposed by \cite{charbonnel07} to reproduce some abundance trends on the red giant branch \citep[e.g.][]{stancliffe08}. In Fig.~\ref{fig:physics}, we show results obtained with our code, using different prescriptions for thermohaline. The differences may reach $0.4$~dex.

We also consider the effect of the initial abundance of metals according to either \cite{asplund09}, as in our computations, or \cite{grevesse93}, as in the previous ones. The resulting differences are small in this case. We note that in this paper, we also carry out computations for very small metallicities ([Fe/H] = $-3.0$, $-4.0$ and $-5.45$), which are smaller than these already studied in the past ([Fe/H] = $-2.31$). We also used smaller accreted masses than before.

\subsection{Back to context}
More than $75\%$ of massive stars in our Galaxy belong to binary or multiple systems and their evolution strongly depends on their interactions with smaller companions \citep{sana12}. Conversely, most of the very old low-mass stars that are still on the main sequence must have been part of binary or multiple systems including more massive companions. These secondary stars have had various opportunities to accrete matter from the winds of the more evolved primaries. As a consequence, the chemical composition of the outer layers of these secondaries was modified according to the given accretion episodes. 

The chemical transformation depends on many parameters. Firstly, the mass of the primary determines the age of the accretion episode and the chemical composition of the wind. Secondly, the distance between both stars at the time of accretion affects the accreted mass. Thirdly, the internal constitution of the secondary depends on its mass and age. Last but not least, the atomic diffusion processes that occurred in the secondary before accretion act as important constraints on the resulting final abundances. When the secondary accretes matter of heavier mean molecular weight than its own, thermohaline convection develops. This mixes the accreted matter downwards. If, however, atomic diffusion had time to build a stabilising $\mu$-gradient below the outer stellar convective zone, before accretion, this gradient acts against thermohaline mixing and prevents it from developing deep inside the star.

Such a scenario has been successfully invoked to account for the carbon-enhanced metal-poor stars enriched in s-process elements (CEMP-s), with winds coming from AGBs onto the proto-CEMP-s \citep[][and references therein]{FuIkIb2000,stancliffe07,stancliffe08,stancliffe09,masseron10,masseron12,MaSt2016,MaSt2017}. This scenario offers a straightforward explanation for the carbon enhancements and also the s-process overabundances as observed in most of these stars.
CEMP-no stars (without any excess of s-process elements) also show a binary fraction of 32\% \citep{arentsen19}. Lithium abundances could then be affected by thermohaline mixing induced by wind accretion in this type of CEMP star too.
Accretion from other types of evolved stars, such as  type II supernovae (SNII), failed SN, white dwarf collapses, non-standard AGBs, have been invoked to account for very metal-poor stars with different chemical compositions, with the possible effect of depleting lithium as compared to that of the plateau . 

\subsection{Lithium in old metal-poor stars}

In any case, if accretion occurs early enough onto the main-sequence companion (where early enough means that the primary star is massive enough), no stabilising $\mu$-gradient had time to build before the accretion episode and thermohaline convection can mix the accreted matter down to the lithium destruction regions. We thus expect, in quite a natural way, for lithium to be depleted in many of these stars as long as the total metallicity of the wind is larger than that of the accreting star. We note that the lithium destruction depends only on the metallicity of the wind, not on the detailed chemical composition. Furthermore, this depletion naturally occurs in various ways, according to the  aforementioned parameters of the stellar system. We thus expect a large lithium dispersion, as has been observed.

Here, we have focused our quantitative study on the CEMP-s case because these stars have been extensively observed and their formation is believed to be caused by a wind accretion from AGB companion. The chemical compositions of these winds are available in the literature \citep{stancliffe08,campbell08}, so it was possible for us to do precise computations on the induced lithium abundances. We found a large dispersion in the final lithium abundances, as observed. The lithium depletion generally occurs in concordance with carbon and heavy element enhancements, except for the cases of massive AGB stars and low-age accretion. In this case, lithium is destroyed but the greater mixing below the convection zone also reduces the accreted overabundance to a point where it is no longer observable.

\subsection{Accreted masses}

The accreted masses needed to deplete lithium are quite small. The minimum is around $0.5~M_{\jupiter}$, which corresponds to about $5\times10^{-4}~M_\odot$. An accreted mass of $4~M_{\jupiter}$, i.e. $4\times10^{-3}~M_\odot$ is enough to reduce the lithium abundance by three orders of magnitude. Some general estimates of the accreted mass on to a stellar companion have been obtained with numerical simulations according to the Bondi--Hoyle--Lyttleton theory, without angular momentum loss \citep{stancliffe07} -- or with angular momentum loss and wind-assisted Roche-lobe overflow \citep{abate15}. The estimates of the mean accreted masses in all these cases lie between 0.002 and 0.35~$M_\odot$. This is larger than what is needed to destroy lithium. The orders of magnitude are however not so different and we must recall that around half of the observed lithium abundances given in these stars are in fact upper limits (Fig. 4 and 5). It is also important to note that CEMP-s with a lithium surface abundance at the plateau value are observed (see dark grey symbols of Fig. \ref{fig:3} and \ref{fig:3bis} for observed lithium abundances that are not upper limits). Our models suggest that such small accreted masses are common and should be considered in future studies of the formation of CEMP-s stars.

\subsection{Chemical content of AGB winds}\label{LiAGB}

In this work, we use the chemical composition of AGB winds from two works:\ one by \cite{stancliffe08} for [Fe/H]$=-2.31$ and the other by \cite{campbell08} for smaller metallicities. These wind compositions were determined with different codes and methods, so that the effect of the metallicity on the resulting surface abundances of CEMP-s stars could be biased when comparing the models at [Fe/H]=-2.31 with the others. Furthermore, the results on the chemical composition of AGB winds at low metallicity are subject to large uncertainties \citep[e.g.][]{campbell08}. In spite of this, we clearly see an impact of the metallicity on the accreted masses needed to deplete lithium. Our results show that exploring the impact of metallicity is crucial for the modelling of CEMP-s stars. 

The wind compositions used in this work do not include the possible increase of lithium due to thermohaline mixing inside the AGB star itself. Without this effect, the lithium abundance in winds lies between 0.15 and 0.4~dex (see A(Li) in Table \ref{tab:1}). According to \cite{stancliffe10}, at [Fe/H]$=-2.31$, the lithium abundance can reach 1.03~dex for a 1~M$_\odot$ AGB, 2.51~dex for a 1.5~M$_\odot$ AGB, and 1.38~dex for a 2.0~M$_\odot$ AGB. No such computations are available at the lower metallicities that we study and report on here. We tried nevertheless to evaluate the possible effects of such a lithium overabundance in the AGB wind. An accreted mass of around 0.5~$M_{\jupiter}$ is needed to start the lithium depletion induced by nuclear reactions This mass is about $12$ times smaller than the mass of the surface region where the chemical elements are homogenised by the turbulent mixing (see Table~\ref{tab:2} for $M_\mathrm{mix}$) in the proto-CEMP-s star models. This implies that the potential lithium overabundance in the wind is diluted in the stellar outer layers just after accretion. Then lithium destruction by nuclear reactions at the bottom of the outer mixed zone should erase, in most cases, the initial abundance differences -- if there are any. Furthermore, the region where the chemical composition is homogenised is even larger when thermohaline convection occurs.

In order to test this effect quantitatively, we computed additional models including a lithium abundance of A(Li)$=2.51$~dex in the wind of a 1 and 2~M$_\odot$ AGB star at [Fe/H]$=-2.31$. We made this choice because it is the maximum lithium abundance predicted by \cite{stancliffe10}. The accreted mass in both models is 0.5~$M_{\jupiter}$. The proto-CEMP-s models have a mass of 0.75~M$_\odot$. For both models, the increase of the lithium abundance in the winds has an impact of at most 0.005~dex on the CEMP-s surface abundance at the turn-off. This comes from the fact that at the moment of accretion, the surface lithium abundance of the proto-CEMP-s model is somewhere between the primordial value ($2.74$~dex) and the plateau value ($2.2\pm0.2$~dex). This is not so different from the $2.51$~dex of the wind. It implies that even for larger accreted masses, the result would be the same. Moreover, the dilution of the accreted material occurs in a region which is at least ten times more massive. Due to the combination of these two effects, changing the lithium content of the wind leads to an almost negligible effect on the final surface abundance of CEMP-s stars. 

\subsection{Conclusion}

We computed CEMP-s models formed from the accretion of an AGB companion wind using the Montr\'eal/Montpellier evolution code. We showed that in most cases, our models give similar results to previous studies that include similar physics but different prescriptions for atomic diffusion and thermohaline convection at [Fe/H]$=-2.31$ \citep[e.g.][and references therein]{MaSt2016}. We also explored smaller metallicities of [Fe/H]$_\mathrm{ini}$=$-3.00$, $-4.00$ and $-5.45$. We demonstrated that accreted masses larger than 0.5~$M_{\jupiter}$ are needed in order to deplete lithium thanks to thermohaline convection mixing, while smaller accreted masses or no accretion lead to stars with lithium plateau surface abundances. We were able to explain the lithium scatter in CEMP-s stars taking into account different accreted masses, initial metallicities, and masses of AGB companions.

As mentioned in Section 6.4, the lithium abundance in the accreted matter has a negligible impact on the surface lithium of the accreting star. We thus expect a similar lithium evolution at lower metallicities in our range of $T_\mathrm{eff}$, as long as the AGB wind is over-metallic compared to the initial metallicity of the AGB progenitor. For [Fe/H]$_\mathrm{ini} < -5.45$, it is possible to predict observations of stars with surface lithium abundances  close to that of the plateau \citep{aguado19b}, under two possible conditions: firstly, for stars which accreted matter late in their evolution, after which a $\mu$-gradient had time to build below the convective zone; and secondly, for stars that accreted matter earlier but in small quantities. Under all other conditions, lithium abundance should be depleted compared to the plateau value.

In conclusion, the lithium abundances observed in old stars may be accounted for in most cases by simple depletion from an original BBN value. This is exactly what is expected if the observed stars accreted high mean molecular weight matter from more evolved companions. These companions may still be visible or the system may have evolved in such a way that they are no longer detectable. All these computations were performed with an initial lithium abundance as deduced from SBBN. The lithium plateau itself is a consequence of the competition between atomic diffusion and mixing processes. The so-called 'lithium problem'
may disappear for accreting stars in the sense that the abundances of old stars are well understood in terms of lithium depletion with an original abundance as obtained from the cosmological background. Other types of accreting metal-poor stars need to be investigated to extend the implications of this result.

\begin{acknowledgements}
This work was supported by FCT/MCTES through the research grants UIDB/04434/2020, UIDP/04434/2020 and PTDC/FIS-AST/30389/2017, and by FEDER - Fundo Europeu de Desenvolvimento Regional through COMPETE2020 - Programa Operacional Competitividade e Internacionalização (grant: POCI-01-0145-FEDER-030389). MD is supported by national funds through FCT in the form of a work contract. We acknowledge financial support from the "Programme National de Physique Stellaire" (PNPS) of the CNRS/INSU co-funded by the CEA and the CNES, France.
\end{acknowledgements}  

\bibliographystyle{aa} 
\bibliography{main.bib} 

\end{document}